\documentclass[a4paper,conference]{IEEEtran}
\usepackage{times}
\usepackage{epsfig}
\usepackage{graphicx}
\usepackage{amsmath}
\usepackage{amssymb}
\usepackage{pifont}
\newcommand{\cmark}{\ding{51}}

\DeclareMathOperator*{\argmin}{arg\,min}
\usepackage[bookmarks=false,colorlinks]{hyperref}
\usepackage[frozencache=true,cachedir=mintedcache]{minted}

\begin{document}

\title{Speckle Image Restoration without Clean Data}

\makeatletter
\newcommand{\linebreakand}{
  \end{@IEEEauthorhalign}
  \hfill\mbox{}\par
  \mbox{}\hfill\begin{@IEEEauthorhalign}
}
\makeatother
\author{
\IEEEauthorblockN{Tsung-Ming Tai}
\IEEEauthorblockA{NVIDIA AI Technology Center\\
ntai@nvidia.com}
\and
\IEEEauthorblockN{Yun-Jie Jhang}
\IEEEauthorblockA{Department of Computer Science \\
and Information Engineering,\\ 
National Taiwan Normal University\\
60547051S@ntnu.edu.tw}
\and
\IEEEauthorblockN{Wen-Jyi Hwang}
\IEEEauthorblockA{Department of Computer Science \\
and Information Engineering,\\ 
National Taiwan Normal University\\
whwang@ntnu.edu.tw}
\linebreakand
\IEEEauthorblockN{Chau-Jern Cheng}
\IEEEauthorblockA{Institute of Electro-Optical Science and Technology,\\ National Taiwan Normal University\\
cjcheng@ntnu.edu.tw}
}

\maketitle

\begin{abstract}
Speckle noise is an inherent disturbance in coherent imaging systems such as digital holography, synthetic aperture radar, optical coherence tomography, or ultrasound systems. These systems usually produce only single observation per view angle of the same interest object, imposing the difficulty to leverage the statistic among observations. We propose a novel image restoration algorithm that can perform speckle noise removal without clean data and does not require multiple noisy observations in the same view angle. Our proposed method can also be applied to the situation without knowing the noise distribution as prior. We demonstrate our method is especially well-suited for spectral images by first validating on the synthetic dataset, and also applied on real-world digital holography samples. The results are superior in both quantitative measurement and visual inspection compared to several widely applied baselines. Our method even shows promising results across different speckle noise strengths, without the clean data needed.
\end{abstract}


\section{Introduction}
\label{sec:introduction}
Speckle noise is an inherent disturbance in coherent imaging systems such as Digital Holography (DH) \cite{Bianco18}, Synthetic Aperture Radar (SAR) \cite{Argenti13}, Optical Coherence Tomography (OCT) \cite{Zaki17}, and Ultrasound systems \cite{Jabarulla18}. Speckle noise is multiplicative and even non-stationary \cite{Cameron13,Xu18}. Conventional linear filtering is then only sub-optimal, producing significant blurring and reducing image contrast. With the proliferation of coherent imaging applications in optical display, medical imaging,  metrology, remote sensing and biology, the desire for speckle noise removal without degrading restored image quality has sparked the development of large varieties of despeckling methods.


\begin{figure*}[tb]
\begin{center}
    \includegraphics[width=1\textwidth]{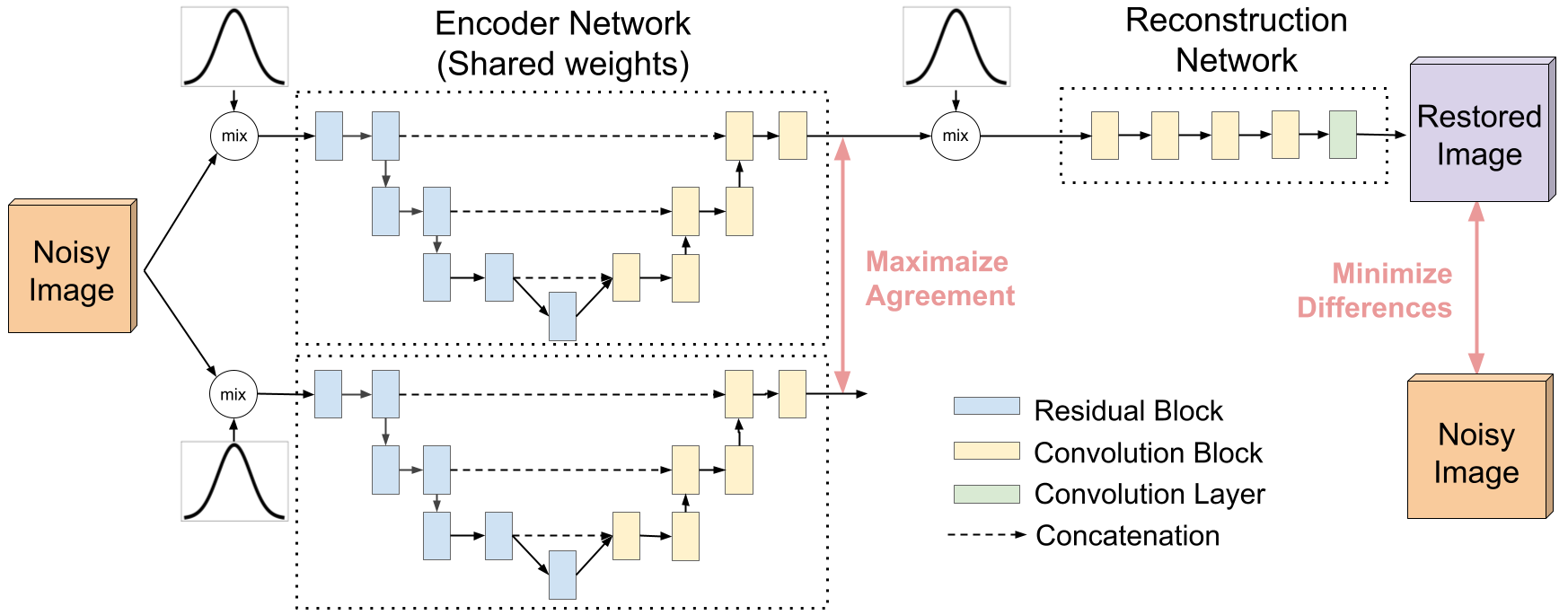}
\caption{The overview of the proposed architecture. Noisy images are firstly augmented by the mixture of noise and fed into the encoder network. The outputs of the encoder network are encouraged to achieve maximal agreements. The encoder output is then transformed by the reconstruction network with an additional mixture of noise and compared with the noisy observations.}
\label{fig:training}
\end{center}
\end{figure*}

Many speckle reduction algorithms are proposed with the prior knowledge of statistical modeling of speckle corruption. For example, algorithms based on Bayesian estimation or linear mean squared estimation \cite{MachineLearningI,Argenti13} can be adopted for speckle noise removal with the density distribution or second-order statistics available. Variants of the algorithms may be beneficial for further performance improvement. These include algorithms carrying out transform domain estimations \cite{Argenti09, Zaki17},  self-similarity operations \cite{Dabov07, Coupe09, Huang19}, or sparse representations \cite{Samuel59}. However, statistical estimations rely on an accurate probabilistic modeling of the signals under concern. The selection of statistical information suitable for modeling the data of interest is usually unavailable. Deep learning techniques are the effective alternatives to statistical modeling approaches for image restoration. Examples of the techniques are the Denoising Auto-Encoders (DAEs) \cite{Vincent08,Bengio13,mao16,Goodfellow}, DnCNNs \cite{zhang2017}, GAN-based denoisers \cite{tripathi2018,chen2018}. However, these methods require noisy observation and clean targets paired during training. 

Some deep learning algorithms can be applied when no clean targets are available. Study in \cite{Ulyanov18} exploits the structure of a neural network to capture image statistics for denoising, thus a single corrupted image may suffice to carry out the noise restoration. However, the whole training process relies on early stopping and is sensitive to the training details. Noise2Noise (N2N, \cite{Lehtinen18}) can denoise without prior knowledge beforehand, but it requires at least a pair of noisy observations for each unknown clean target.  For some coherent imaging systems, the noisy image pairs for unknown clean targets are not available, which makes the N2N inapplicable on these systems. For Noise2Void (N2V) and its variants \cite{Krull19,Krull20}, image restoration is carried out on the local \textit{blind-spot} patches. The central pixel of blind-spot is masked and left as the prediction task for the network to solve. However, blind-spot algorithm assumes the noisy corruption is with zero mean and also the central pixel in the patches are predictable by its surrounding context, which is probably not held in coherent imaging systems. Similar approaches can be also seen in \cite{Xu19,batson2019,lin2019,cha2019}. In addition, denoisers based on self-supervised learning have also been proposed in recent years \cite{laine2019high,fadnavis2020patch2self,quan2020self2self}. However, these denoisers are under the assumption about the noise is conditionally independent to the signal, which greatly degrades their performance when applied to coherent systems with multiplicative signal dependent noise.

The objective of this study is to present a novel deep learning algorithm which is suitable for multiplicative noise removal without clean data. Apart from previous works, we focus on the speckle noise problem in coherent systems, where the noise corruption depends on the pixel intensities and tends to be more severe when the signal is with higher responses. Therefore, the general denoisers which leverage the content-independent noise statistics in prior works cannot effectively remove the speckle noise. We tackle this problem by learning with a robust representation based on agreement maximization. Furthermore, no prior statistic of speckle noise is required in the proposed method. Our modeling also involved a reconstruction network which is able to be weakly supervised from noisy observations.

\section{Algorithm}
\label{sec:algorithm}

\subsection{Preliminary}
Let $\{ {\bf s}^1,...,{\bf s}^N \}$ and  $\{ {\bf y}^1,...,{\bf y}^N \}$ be the set of clean images and noisy observations, where $N$ is the number of images. The additive noise can be described by the following.
\begin{equation}
{\bf y}^j = {\bf s}^j + {\bf a}^j,   \label{eq:additive}
\end{equation}
where ${\bf a}^j$ is an additive noise under independent and identical distributions (i.i.d). 

Given a denoising neural network $A_{\theta}$ parameterized by $\theta$, the goal of the denoising in the supervised  learning is to carry out the below empirical minimization task:
\begin{equation}
\argmin_{\theta} \frac{1}{N} \sum_{j=1}^N {\mathcal L}(A_{\theta}({\bf y}^j),{\bf s}^j),     
\label{eq:cost1}
\end{equation}
where ${\mathcal L}$ can be an arbitrary differentiable distance measurement function, for example, the Mean Squared Error (MSE). It can be observed that the set of clean images $\{ {\bf s}^1,...,{\bf s}^N \}$ is required for training in this case.

Noise2Noise (N2N, \cite{Lehtinen18}) addresses the lack of clean targets with the assumption that the additive noise in~\eqref{eq:additive} has zero-mean. Therefore, the averaging over noisy observations can be approximated to clean targets. Given ${\bf y}'^{j}$ the noisy observation, which shared the same clean context ${\bf s}^j$ as with ${\bf y}^j$, affect by another additive noise instance ${\bf a}'^j$. Then, the following minimization task is able to approximate \eqref{eq:cost1}, without clean targets needed. 
\begin{equation}
\argmin_{\theta} \frac{1}{N} \sum_{j=1}^N {\mathcal L}(A_{\theta}({\bf y}^j),{\bf y}'^j)     
\label{eq:cost2}
\end{equation}

Noise2Void (N2V, \cite{Krull19}) can further be applied when neither paired noisy observation nor clean target are available. Instead of using the whole image as inputs, the patch ${\bf r}^j_i$ of ${\bf y}^j$ is leveraged. The overall objective function for the learning problem is like:
\begin{equation}
\argmin_{\theta} \frac{1}{NP} \sum_{j=1}^N \sum_{i=1}^P {\cal L}(A_{\theta}({\bf r}^j_{i}), y^j_i),     
\label{eq:cost3}
\end{equation}
where the $P$ is the number of total patches generated from ${\bf y}^j$.

Even though N2V has no limitation on the paired noisy observations, it still targets the zero-mean additive noise assumption.

\subsection{Proposed Method} \label{sec:dae}
Following shows the speckle noise model, where the multiplicative noise  ${\bf m}^j$ is applied on each individual pixel location of an image. The generic additive noise is considered in ${\bf a}^j$.
\begin{equation}
{\bf y}^j = {\bf m}^j {\bf s}^j + {\bf a}^j   \label{eq:multiplicative}
\end{equation}

Note we do not explicitly model the distribution of ${\bf m}^j$, as it is usually unknown and may vary from different optical systems. These systems usually produce only a single observation per view angle of the same interest object, which makes N2N inapplicable. Different from additive noise, the speckle noise is pixel intensity dependent which further degrades the effectiveness of patch-based denoisers. On the other hand, our method does not require the paired noisy observations and any clean target as well. The proposed model processes the whole image inputs, and is composed of encoder and reconstruction networks. 

\subsubsection{Encoder with noise mixture} 
Given a noisy observation ${\bf y}^j$ corrupted by speckle noise, we first augment it with the mixture of adaptive noise by:
\begin{equation}
{\bf \widehat{y}}^j =  {\bf \widehat{m}}^j {\bf y}^j  + {\bf \widehat{a}}^j \label{eq:step1}
\end{equation}
where ${\bf \widehat{m}}^j$ and ${\bf \widehat{a}}^j$ are drawn from the Gaussian distribution centered at 1 and 0 for multiplicative and additive parts respectively,  with the variance, $\sigma_y^j$, adapted by inputs ${\bf y}^j$. In details, ${\bf \widehat{m}}^j \sim {\mathcal N}(1, \sigma_y^j)$ and ${\bf \widehat{a}}^j \sim {\mathcal N}(0, \sigma_y^j)$.

The input are augmented with two different noise instances using the same process in~\eqref{eq:step1}, results in ${\bf \widehat{y}}^{j,1}$ and ${\bf \widehat{y}}^{j,2}$, then being fed to the encoder network, $f_\theta$. The outputs are regularized by the agreement maximization loss:
\begin{eqnarray}
{\bf z}^{j,k} &=& f_\theta({\bf \widehat{y}}^{j,k}),~~~\text{for $k=1,2$} \label{eq:encoder} \\
\mathcal{L}_{\text{agreement}} &=& \frac{1}{N} \sum_{j=1}^N MSE({\bf z}^{j,1}, {\bf z}^{j,2})  \label{eq:c1}
\end{eqnarray}

The agreement maximization is expected to extract noise-free representation. The intuition is that the augmented inputs are equivalent to forming the in-model paired observations and be able to approximate the clean target, which is similar to \cite{Lehtinen18}.  It can be seen by combining \eqref{eq:multiplicative}\eqref{eq:step1},
\begin{equation}
\begin{aligned}
{\bf \widehat{y}}^{j,k} &= {\bf \widehat{m}}^{j,k} ({\bf m}^j {\bf s}^j + {\bf a}^j) + {\bf \widehat{a}}^{j,k} \\
&= {\bf m}^{j,k}_{\text{aug}} {\bf s}^j + {\bf a}^{j,k}_{\text{aug}},~~~\text{for $k=1,2$}
\label{eq:syn}
\end{aligned}
\end{equation}
where ${\bf m}^{j,k}_{\text{aug}}$ and ${\bf a}^{j,k}_{\text{aug}}$ are the augmented noise. For each training iteration, they are re-sampled using the proposed adaptive noise mixture to further produce the sufficient in-model paired observations for approximation to the noise-free latent representations.

\subsubsection{Reconstruction with noise mixture}
We further reconstruct the encoder outputs back to images from latent representations. By doing so, a reconstruction network $g_\phi$ is deployed. To avoid the supervision of noisy observation bringing back the noise and further polluting the encoder output, an additional noise mixture is applied, as the same as \eqref{eq:step1}, on the latent input of the reconstruction network. 
\begin{eqnarray}
{\widehat{\bf z}}^j &=& {\bf \widehat{m}}^j {\bf z}^{j,1} + {\bf \widehat{a}}^j \label{eq:reconstruct_noise} \\
{\bf \widehat{s}}^j &=& g_\phi({\widehat{\bf z}}^j) \label{eq:reconstruct}
\end{eqnarray}

This noise injection step, in the backward viewpoint, acts to jitter the uncertainty to the gradient signals. Note we only take a single branch of the encoder output in~\eqref{eq:encoder}. The ${\bf \widehat{s}}^j$ is the estimation of the clean target in the proposed algorithm. The reconstruction loss is to minimize the differences between target estimation and noisy observation.
\begin{equation}
\mathcal{L}_{\text{reconst}} = \frac{1}{N} \sum_{j=1}^N MSE({\bf \widehat{s}}^j, {\bf y}^j)  \label{eq:c2}
\end{equation}

The overall loss function is the sum of $\mathcal{L}_{\text{agreement}}$ in \eqref{eq:c1} and $\mathcal{L}_{\text{reconst}}$ in \eqref{eq:c2}. They are optimized jointly and can be trained end-to-end.
\begin{equation}
\mathcal{L}_{\text{overall}} = \mathcal{L}_{\text{agreement}} + \mathcal{L}_{\text{reconst}}  \label{eq:final_loss}
\end{equation}


Note that the mixture of noises are only required during training, they are not needed and disabled in inference. 

\begin{figure}[t]
\center
\includegraphics[width=0.65\columnwidth]{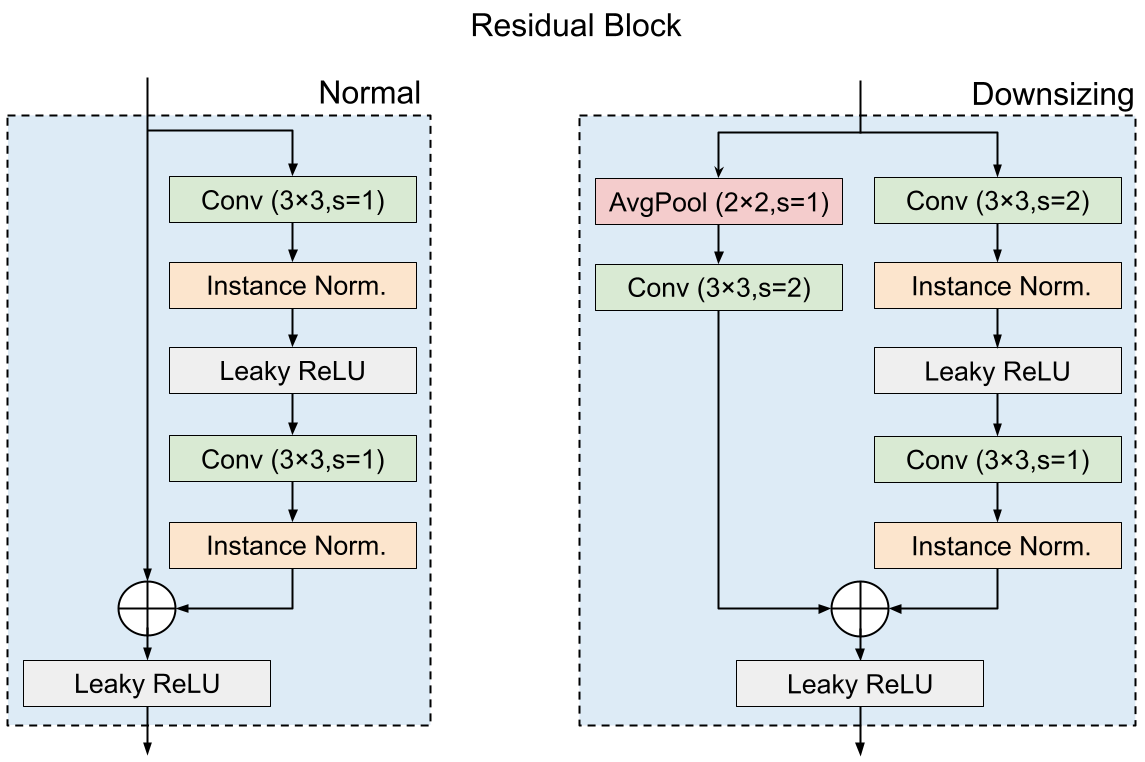}
\includegraphics[width=0.25\columnwidth]{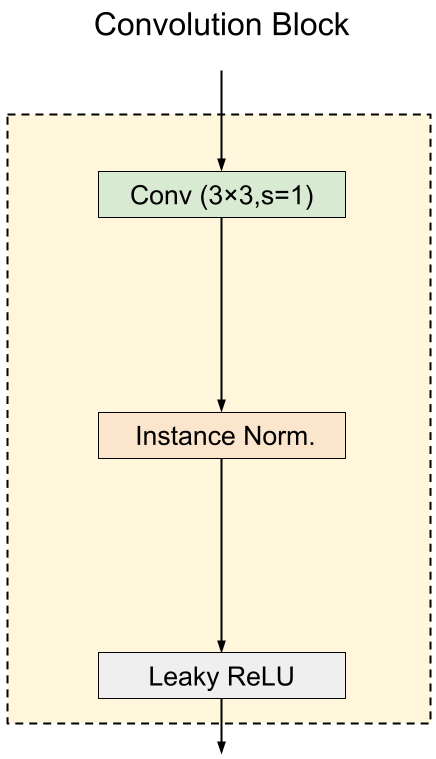}
\caption{Details of ResidualBlocks and ConvBlocks used in our model.}
\label{fig:block}
\end{figure}

\begin{table}[t]
\caption{Architecture details of proposed model.}
\center
\scriptsize
\begin{tabular}{ll|c|l|l}
\hline
Component & Name & \#Blocks & Shape & \#Filters \\
\hline
& Noise Mixture & 1 & 256$\times$256 & 1 \\
& Stem (ConvBlocks) & 1 & 256$\times$256 & 32 \\
& ResidualBlocks & 1 & 256$\times$256 & 32 \\
& ResidualBlocks & 2 & 128$\times$128 & 64 \\
& ResidualBlocks & 2 &64$\times$64 & 128 \\
& ResidualBlocks & 2 &32$\times$32 & 256 \\
ResUNet & Upsample\&Concat & - & 64$\times$64 & 256+128 \\
(Encoder) & ConvBlocks & 2 & 64$\times$64 & 128 \\
& Upsample\&Concat & - & 128$\times$128 & 128+64 \\
& ConvBlocks & 2 & 128$\times$128 & 64 \\
& Upsample\&Concat & - & 256$\times$256 & 64+32 \\
& ConvBlocks & 2 & 256$\times$256 & 32 \\
& InstanceNorm & 1 & 256$\times$256 & 32 \\
\hline
Reconstruction & Noise Mixture & 1 & 256$\times$256 & 32 \\
Network & ConvBlocks & 4 & 256$\times$256 & 32 \\
& Conv & 1 & 256$\times$256 & 1 \\
\hline
\multicolumn{5}{|c|}{Total 17.35 GFLOPs.} \\
\hline
\end{tabular}
\label{tab:model}
\end{table}

\section{Experimental Results}
\label{sec:exp}
\begin{table*}[!t]
\caption{PSNR values of various denoisers for noisy observations corrupted by speckle noise. The ground truth of the observations are from the COIL-100 dataset. The best numbers are colored in {\color{red}red} and the second best numbers are in {\color{teal}green}.}
\label{tab:coil}
\setlength{\tabcolsep}{1.8\tabcolsep}
\center
\small
\begin{tabular}{l|c|cccccc}
\hline
Denoiser & Need Noise Prior? & Fruits & Cup   & Mug   & Trunk & Tool  & Container \\
\hline
Noisy & & 22.19 dB & 19.42 dB & 19.42 dB & 23.53 dB & 19.96 dB & 19.65 dB \\
\hline
BM3D \cite{Dabov07} & \cmark & 40.13 dB & 37.00 dB & 32.09 dB & 35.13 dB & {\color{red}36.63 dB} & {\color{teal}36.31} dB \\
N2V  \cite{Krull19} & & 33.30 dB & 32.55 dB & 29.12 dB & 24.76 dB & 31.53 dB & 31.34 dB \\
NAC \cite{Xu19} & \cmark & {\color{teal}40.38 dB} & {\color{teal}37.26 dB} & {\color{red}36.21 dB} & {\color{teal}35.26 dB} & 34.80 dB & {\color{teal}36.31 dB} \\
\hline
Ours & & {\color{red}40.67 dB} & {\color{red}40.13 dB} & {\color{teal}34.94 dB} & {\color{red}36.71 dB} & {\color{teal}36.33 dB} & {\color{red}41.15 dB} \\
\hline
\end{tabular}
\end{table*}

\begin{figure*}[!t]
\begin{center}
\includegraphics[width=1\textwidth]{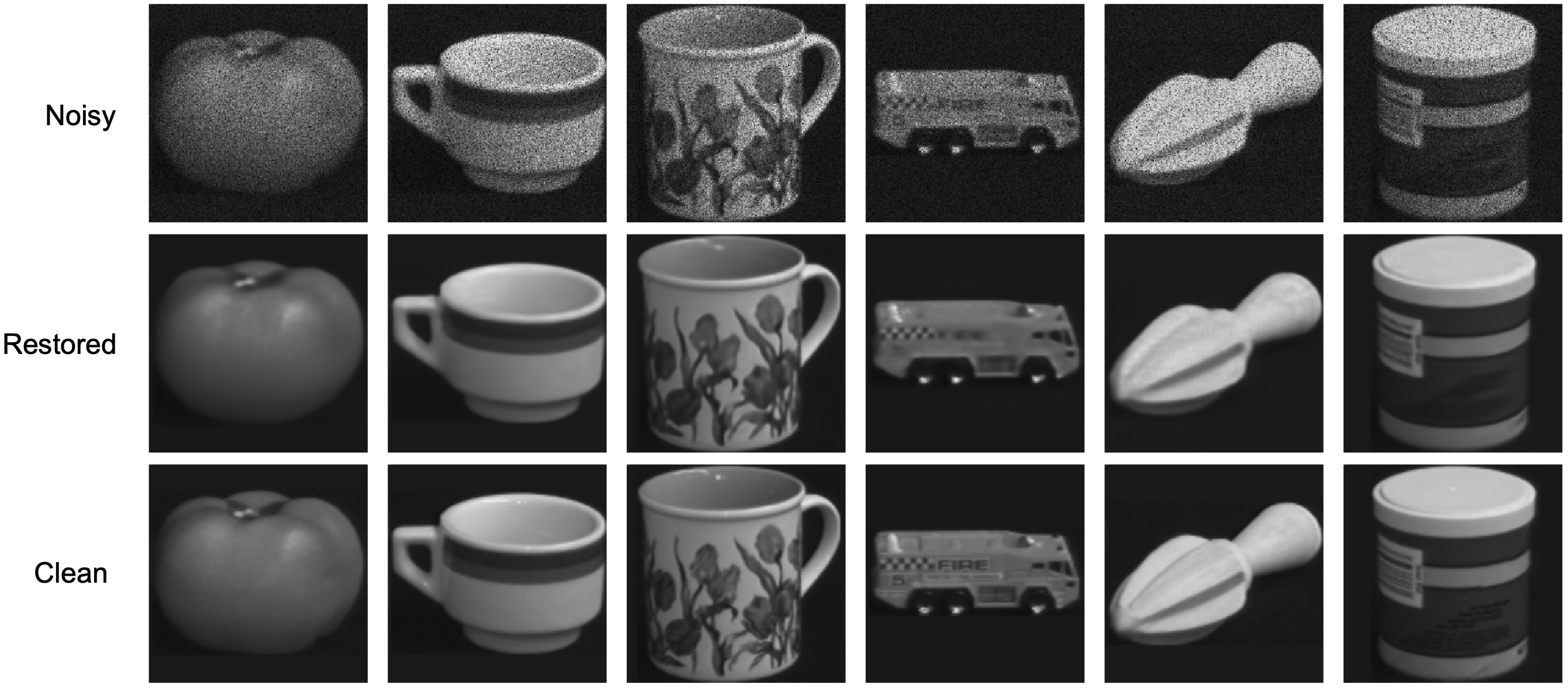}
\caption{The noisy, restored, and clean images of COIL-100 objects (\textit{Fruits}, \textit{Cup}, \textit{Mug}, \textit{Trunk}, \textit{Tool}, \textit{Container} are visualized. The noisy images are acquired by the mixture of speckle noise with the clean images. The restored images are the results produced by the proposed algorithm.}
\label{fig:coil}
\end{center}
\end{figure*}


\subsection{Implementation Details}
We adopt ResUNet \cite{zhang2018} as the encoder network, followed by a sequence of convolution blocks for the reconstruction network, shown in Figure~\ref{fig:block} and Table~\ref{tab:model}. Experiments are conducted on a single NVIDIA RTX 3090 GPU. AdamW \cite{loshchilov18} optimizer is adopted, with a linear learning rate annealing from 3e-3 downto 0. Weight decay is set to 5e-4. We also applied gradient norm clipping with 1e-2. Images are resized to 256 x 256 for all datasets.  We set the total training epochs to 500 for synthesized images, 200 for digital holograms.

\subsection{Datasets}
\subsubsection{Synthesized Datasets}
\label{sec:synthesized_dataset}
The subset of COIL-100 \cite{nene1996} and F-16 images are considered in the synthesized datasets. For COIL-100, we consider six different representative objects: shape and shadow (\textit{Fruits}, \textit{Tool}), text (\textit{Container}), and complex patterns (\textit{Mug}, \textit{Trunk}). Each object contains 72 images in different view angles of the object. All the synthesized images in the experiments obey the following speckle noise model,
\begin{equation}
{\bf y}^j = \alpha {\bf m}^j {\bf s}^j
\label{eq:noise_model}
\end{equation}
where ${\bf m}^j$ is the multiplicative speckle noise applied on the clean images, $s^j$. We make ${\bf m}^j$ drawn from the Gaussian distribution centered at 1. By adjusting $\alpha$ ($\alpha > 0$), we can control the noise strength levels of synthesized speckle noise. For all the experiments using the synthesized dataset, we trained on 80\% of samples and left the final 20\% for testing.

\subsubsection{Real-World Digital Holography}
The digital holograms are obtained by an off-axis optical setup of a modified Mach–Zehnder interferometric architecture with a CCD for recording light interference patterns of an object \cite{chen2019}. Subsequent numerical operations are then carried out to obtain the reconstructed images of the object from digital holograms. The ground truth for the noisy digital hologram reconstruction is unknown. Each object is captured in 360 images for different view angles (288 for training and 72 for testing). Note that the underlying noise distribution for the multiplicative part is unknown; these real-world samples may also come with the additive noise together with the speckle noise.

\begin{figure}[!t]
\begin{center}
\includegraphics[width=1\columnwidth]{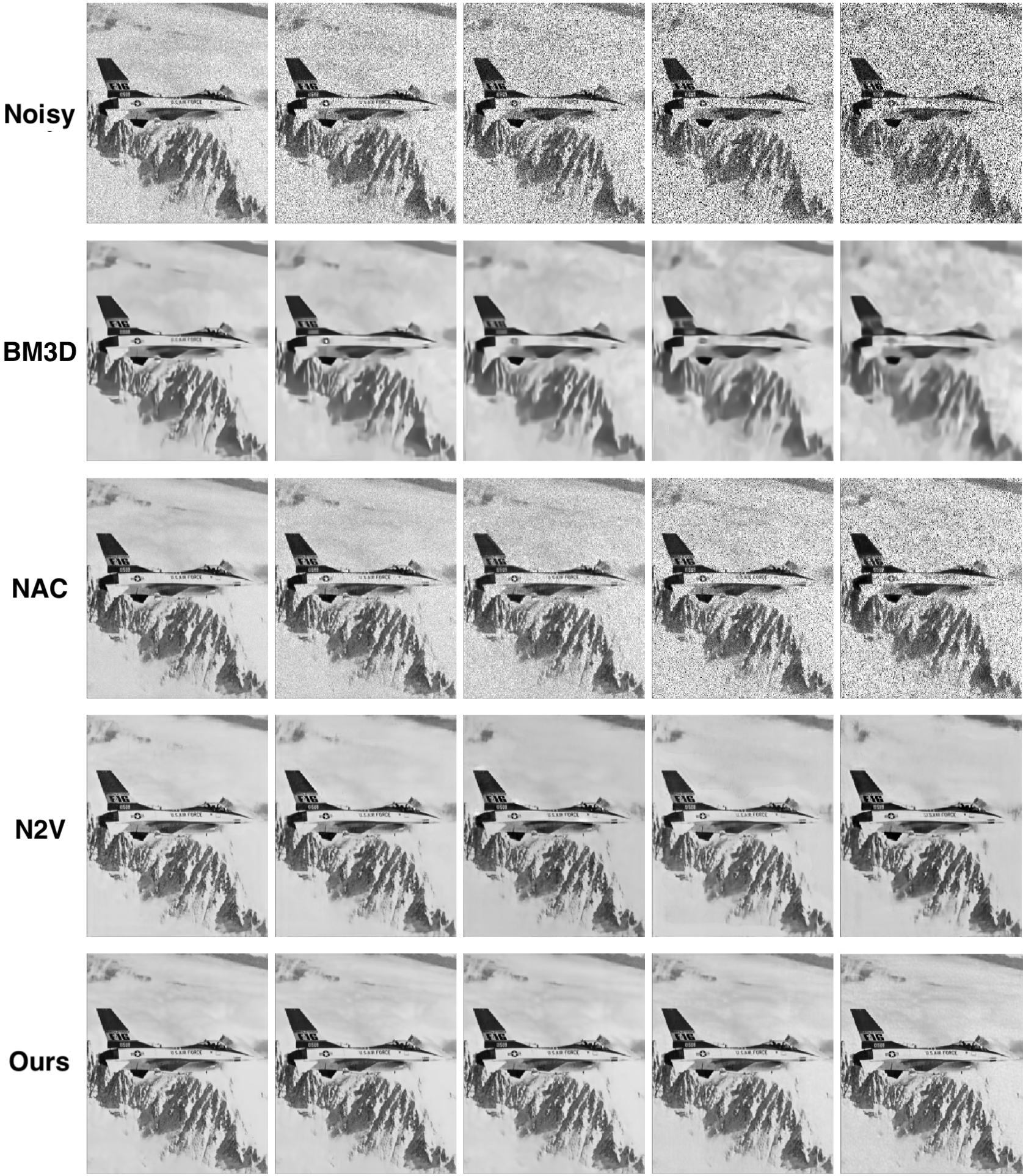}
\caption{The visualization of restored F-16 images for various denoisers. The different noise levels are presented in the column order, from left to right are $\alpha = [0.1, 0.2, 0.3, 0.4, 0.5] \pm 0.05$. Noise corrupted inputs and the restored images for different denoisers are presented in rows from top to bottom.}
\label{fig:f16}
\end{center}
\end{figure}

\begin{table*}[!t]
\caption{PSNR measured from various denoisers at different noise levels in F-16 synthesized images.}
\setlength{\tabcolsep}{1.8\tabcolsep}
\label{fig:f16_plot}
\begin{center}
\begin{small}
\begin{tabular}{lccccc}
\hline
Method & $\alpha=0.1$ & $\alpha=0.2$ & $\alpha=0.3$ & $\alpha=0.4$ & $\alpha=0.5$   \\
\hline
Noisy & 22.14 dB & 17.13 dB & 16.21 dB & 12.69 dB & 11.17 dB \\
BM3D \cite{Dabov07} & 29.51 dB & 28.19 dB & 28.04 dB & 27.62 dB & 27.41 dB \\
N2V \cite{Zaki17} & 31.46 dB & 31.26 dB & 30.70 dB & 30.85 dB & 30.72 dB \\
NAC \cite{Xu19} & 35.94 dB & 28.82 dB & 26.97 dB & 24.38 dB & 23.26 dB \\
\hline
Ours (w/o reconstruct noise mixture) & \textbf{48.25 dB} & 38.09 dB & 33.55 dB & 29.81 dB & 23.49 dB \\
Ours & 41.19 dB & \textbf{38.83 dB} & \textbf{38.07 dB} & \textbf{38.64 dB} & \textbf{38.42 dB} \\
\hline
\end{tabular}
\end{small}
\end{center}
\end{table*}

\begin{figure}[t]
\begin{center}
\includegraphics[width=1\columnwidth]{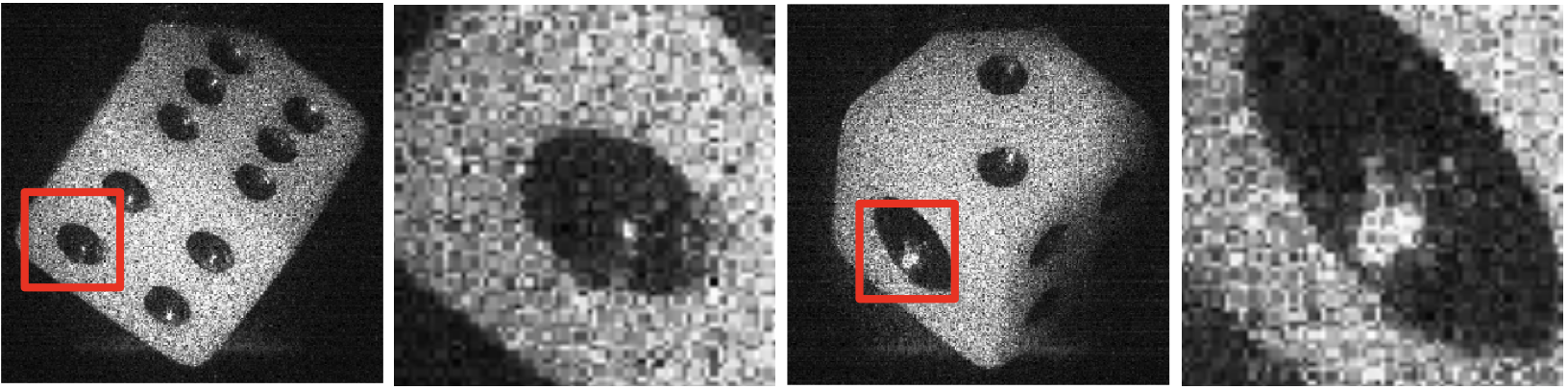}
\includegraphics[width=1\columnwidth]{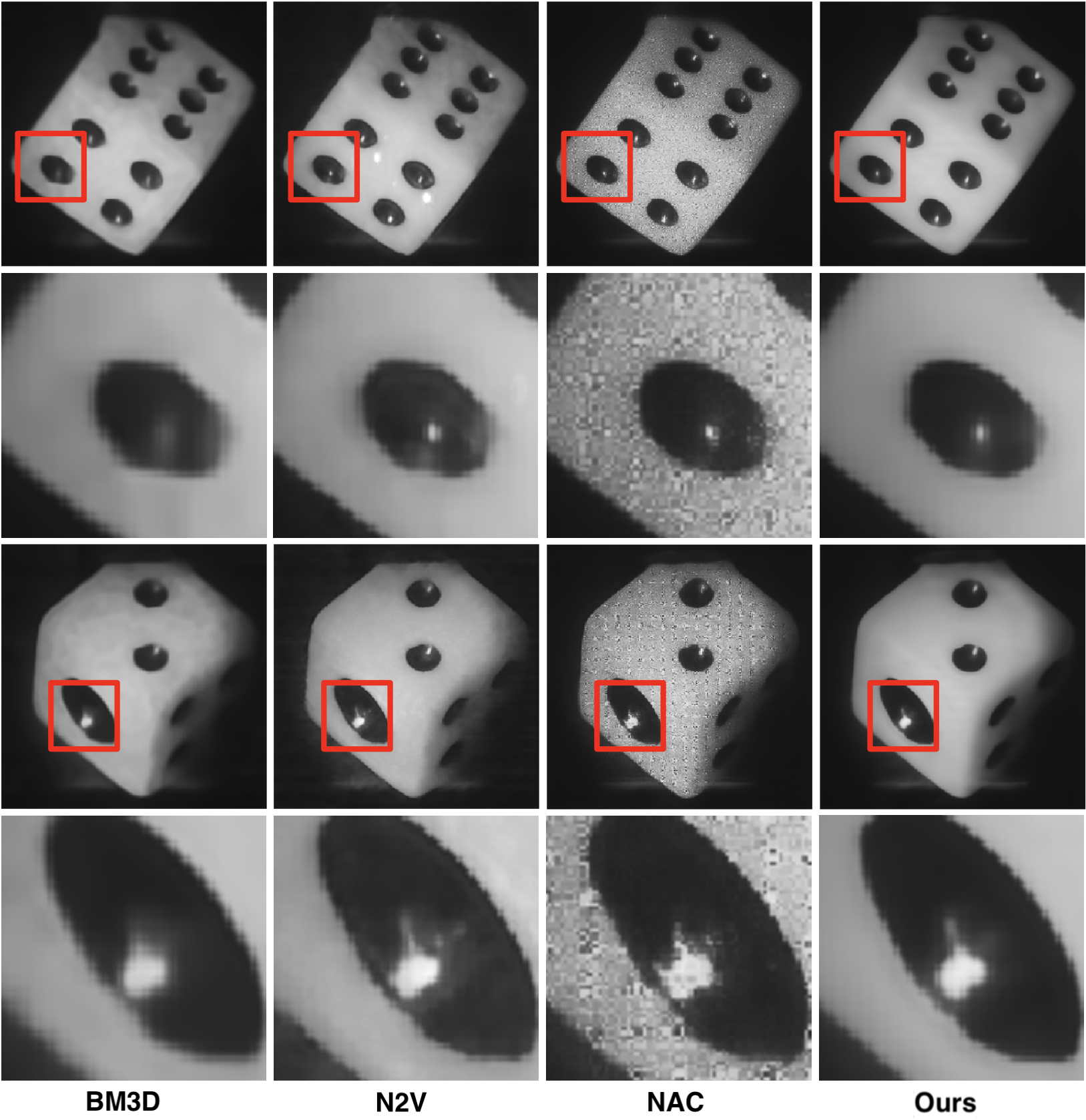}
\caption{Image restoration results on digital holograms. The first row presents two reconstructed digital holograms containing speckle noise. Each shows different sides of a die. The rest of the rows show restored results from different denoisers, where the first two rows presented one side of the die and followed by another side.}
\label{fig:dice}
\end{center}
\end{figure}

\subsection{Results}
For noise synthesized datasets, we measure the Peak Signal Noise Ratio (PSNR) between the restored and ground-truth images. Table~\ref{tab:coil} summarizes the numerical results on the COIL-100 dataset, with noise strength set to $\alpha=0.3$. We compare the proposed method with different denoisers when only a single noisy observation for each unique viewpoint is available. BM3D and Noise-As-Clean (NAC) are the strong baselines for speckle removal, but the prior information (i.e., estimated noise strength) needs to be known beforehand. Noise2Void (N2V) , relying on the local patches and additive Gaussian noise assumption, shows degraded performance on speckle noise. The proposed method without using any noise prior shows overall better performance than other baselines.

Table~\ref{fig:f16_plot} shows the PSNR comparison in different noise levels on F-16 images. We synthesized a total of 500 noisy observations for $\alpha \in \{0.1, 0.2, 0.3, 0.4, 0.5\}$ with variance $\pm0.05$.  Our method has superior PSNR values among all the other methods for all the noise levels. We also include a design variant by removing the noise mixture on the input of the reconstruction network (${\bf \widehat{m}}^j$ and ${\bf \widehat{a}}^j$ in \eqref{eq:reconstruct_noise}). It shows the results when the noise level increases, the PSNR without the noise mixture in the reconstruction network degrades rapidly. This suggests that a further mixture of noise in the reconstruction network is needed. Figure~\ref{fig:f16} demonstrates the restored images produced from each denoiser visually. Our algorithm shows the best visual quality on the restored images for all the noise levels. Even under the strongest noise level (e.g., $\alpha=0.5$), the image details are still well-preserved. By contrast, artifacts or severe noise corruptions are observed in the other methods. 
\begin{table*}[!t]
\caption{PSNR measurements when different numbers of training samples are used.}
\setlength{\tabcolsep}{1.8\tabcolsep}
\label{fig:ablation1}
\begin{center}
\begin{small}
\begin{tabular}{lcccccc}
\hline
Method & 15 samples & 30 samples & 60 samples & 125 samples & 250 samples & 500 samples \\
\hline
Noisy & 17.13 dB & 17.13 dB & 17.13 dB & 17.13 dB & 17.13 dB & 17.13 dB \\
BM3D \cite{Dabov07} & 28.19 dB & 28.19 dB & 28.19 dB & 28.19 dB & 28.19 dB & 28.19 dB \\
N2V \cite{Zaki17} & 28.30 dB & 29.20 dB & 30.70 dB & 31.03 dB & 31.14 dB & 31.26 dB\\
NAC \cite{Xu19} & 25.94 dB & 26.11 dB & 26.29 dB & 28.48 dB & 28.66 dB & 28.82 dB\\
Ours & \textbf{32.62 dB} & \textbf{34.59 dB} & \textbf{35.97 dB} & \textbf{38.54 dB} & \textbf{38.72 dB} & \textbf{38.83 dB} \\
\hline
\end{tabular}
\end{small}
\end{center}
\end{table*}

\begin{table}[t]
\caption{PSNR measurement of individual design component}
\setlength{\tabcolsep}{1.5\tabcolsep}
\label{tab:ablation_effectiveness}
\center
\small
\begin{tabular}{lc}
\hline
Components & PSNR \\
\hline
Base Architecture & 17.15 dB \\
+ Reconstruction noise mixture & 18.52 dB\\
+ Encoder noise mixture and agreement loss & 38.09 dB\\
+ Both (Proposed Method) & \textbf{38.83 dB}\\
\hline
\end{tabular}
\end{table}

\begin{table}[t]
\caption{The variances of feature distance vector \textbf{d}.}
\setlength{\tabcolsep}{1.8\tabcolsep}
\label{tab:ablation3}
\center
\small
\begin{tabular}{lc}
\hline
Method & Variance of \textbf{d} \\
\hline
N2V ($\alpha=0.2$) & 20.66 \\
NAC ($\alpha=0.2$) & 58.16 \\
\hline
Ours ($\alpha=0.1$) & \textbf{1.68} \\
Ours ($\alpha=0.2$) & \textbf{1.75} \\
Ours ($\alpha=0.3$) & \textbf{1.52} \\
Ours ($\alpha=0.4$) & \textbf{1.45} \\
Ours ($\alpha=0.5$) & \textbf{1.56} \\
\hline
\end{tabular}
\end{table}

\subsection{Denoising Real-World Corrupted Digital Holograms}
\label{sec:dice_exp}
The digital holograms with speckle noise corruption are shown in Figure~\ref{fig:dice}. Note both ground-truth of clean source images and the noise statistics are not available. The evaluations are therefore presented by visual inspection. The first row in the figure shows the original digital holography images corrupted by speckle noise. We zoom in and highlight two local areas for detailed inspection. Row 2-4 shows the restored images with each column presented by different denoising algorithms. We tune for BM3D and NAC of required noise strength by a grid search to obtain the best visual quality. The proposed method is able to effectively remove the speckle noise with minimal artifacts without any further tuning. In addition, artifacts and noisy corruptions can still be identified from the de-speckled images produced by other methods. Overall, our method has significantly superior visual quality and doesn't require any noise statistical information.

\subsection{Samples Efficiency}
Table~\ref{fig:ablation1} shows the PSNR measurements of F-16 images in $\alpha = 0.2$ with different methods when training on 15 to 500 samples. In the case noisy observations are few (i.e., 15), we are 4.32 db better than the second best number in the table. The proposed model also shows increasing performance while the observations increase, and achieves 7.57 dB better than the second candidate in a total of 500 training samples. The gap shows our proposed method not only can utilize few noisy observations efficiently, but also enjoys performance gains when more observations are available.

\subsection{Design Ablations}
Table~\ref{tab:ablation_effectiveness} validates the effectiveness of each proposed design of our model. The experimental results are conducted on synthesized F-16 images with noise level $\alpha=0.2$. With no dual-path and any mixture of noise, identity function is learned and shows no effect on the denoising task. When the reconstruction network with a noise mixture is deployed, the learning is more resilient to the noise in the supervision target and shows marginal improvement. By deploying the dual-path encoder with individual noise mixture and agreement loss, a significant improvement from 17.15 dB to 38.09 dB is observed. The complete modeling is coming from both designs, further improved from 38.09 dB to 38.83 dB.

\subsection{Latent Robustness Analysis}
Considering the encoder output is in the high dimensional space, a numeric analysis is performed to validate the effectiveness of proposed agreement loss by measuring the latent distance. We define the distance vector ${\bf d}=\{{\bf{d}}^1, ... {\bf{d}}^N\}$ of the encoder output ${\bf z}^j$ in~\eqref{eq:encoder} as follows,
\begin{eqnarray}
    {\bf z}^{ref} &=& \frac{1}{N} \sum^N_j {\bf z}^j, \\
    {\bf d}^j &=& \sqrt{ \frac{1}{D} \sum_i^D{ ({\bf{z}}_i^j - {\bf{z}}_i^{ref})^2 } },
\end{eqnarray}
where $D$ is the feature dimension of the encoder output. $N$ is the total number of the samples. We record the variance of ${\bf d}$ in total $N=500$ at F-16 noisy observations in different speckle noise strengths. Our method consistently shows low variances within each noise level in Table~\ref{tab:ablation3}, and remains constant across different noise levels. This result aligns with our hypothesis that the encoder output regularized by agreement loss can approximate a noise-free representation of a clean target. Otherwise, the variances should be proportional to the input noise levels. In contrast, other methods that rely on other denoising mechanisms obtain higher variances. This illustrates the underlying philosophy is different in our design.

\section{Conclusion}
In this paper we presented a novel speckle noise removal method, by mixtures of adaptive noises in both encoder agreement and reconstruction network. The proposed method can be applied when no clean target and noise statistics are available. Unlike previous works, we aim to form a robust representation resilient to noise corruptions instead of estimating the noise model directly. In both synthesized and real-world holography scenarios, the proposed method shows significantly better visual quality and numeric evaluation compared with different widely adopted denoisers.

\bibliographystyle{IEEEtran}
\bibliography{main}

\begin{figure*}
    \centering
    \includegraphics[width=1\textwidth]{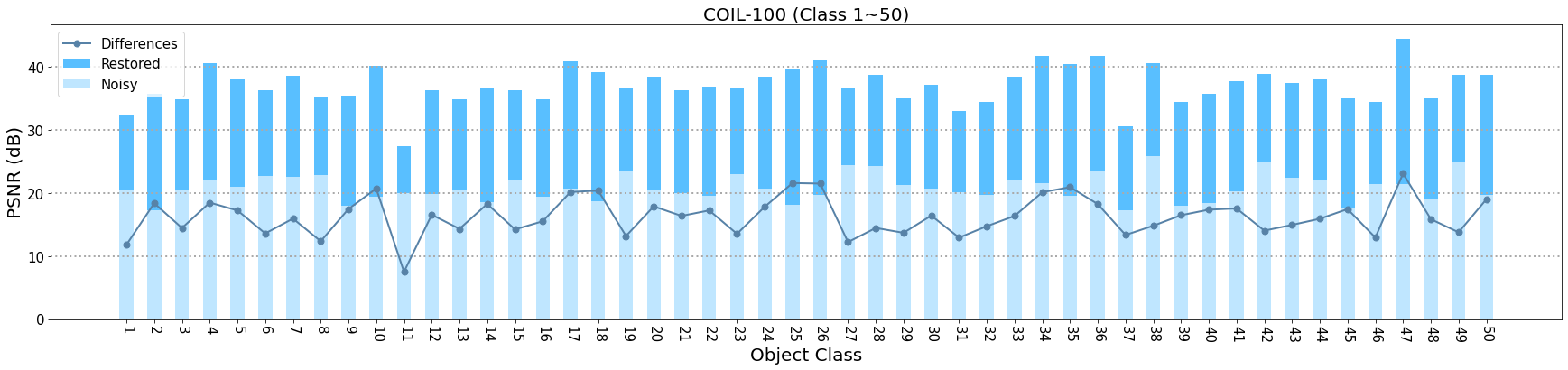}
    \includegraphics[width=1\textwidth]{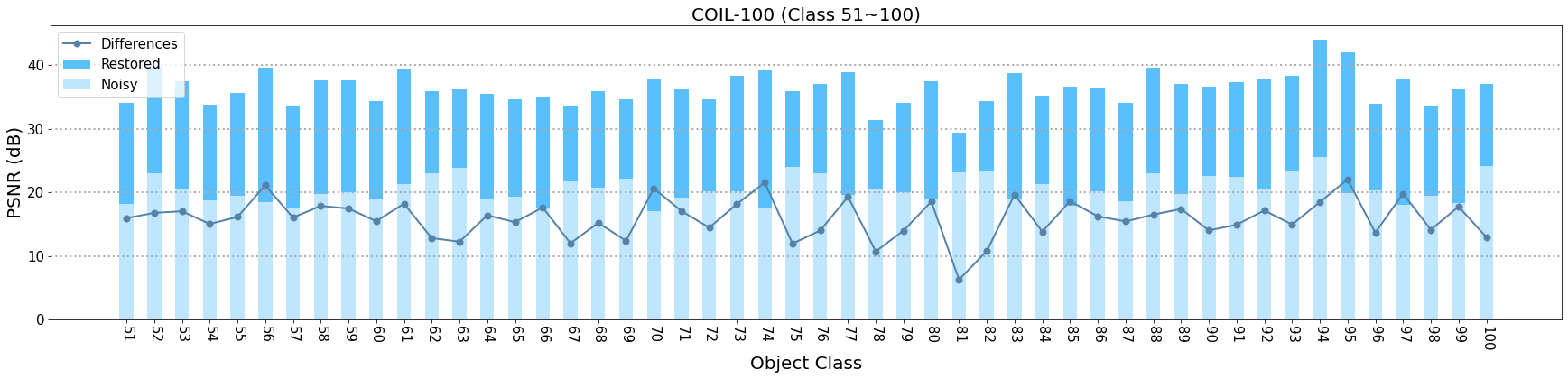}
    \caption{PSNR evaluations of all classes in the COIL-100 dataset. The PSNR measurements between noisy observations and restored images are computed. The differences are also shown by the line plot.}
    \label{fig:all_coil100_test}
\end{figure*}

\newpage
\appendix
\section{Complete evaluation on COIL-100 dataset.}
In Figure~\ref{fig:all_coil100_test}, we provided the complete evaluation over the all 100 classes. Our proposed method can restore the noise corruptions from averaged 20.68 PSNR of noisy observations, improving to averaged 36.77 PSNR, which is +16.09 PSNR improvement.

\end{document}